\documentstyle[psfig]{elsart}

\begin{document}
\newcommand{\beq}[1]{\begin{equation}\label{#1}}
\newcommand{\eeq}{\end{equation}}
\newcommand{\beqar}[1]{\begin{eqnarray}\label{#1}}
\newcommand{\eeqar}{\end{eqnarray}}
\newcommand{\lash}[1]{\not\! #1 \,}
\newcommand{\bra}[1]{\big< #1 \big|}
\newcommand{\ket}[1]{\big| #1 \big>}
\newcommand{\PR}{Phys. Rev.\ }
\newcommand{\PRL}{Phys. Rev. Lett.\ }
\newcommand{\PL}{Phys. Lett.\ }
\newcommand{\NP}{Nucl. Phys.\ }
\newcommand{\ZP}{Z. Phys.\ }
\newcommand{\AP}{Ann. Phys.}
\begin{frontmatter}
\title{Single spin asymmetry for
the Drell-Yan process}

\author{$^1$N. Hammon}, 
\author{$^2$O. Teryaev},
\author{$^1$A. Sch\"afer}
\address{$^1$Institut f\"ur Theoretische Physik
Johann Wolfgang Goethe-Universit\"at, Frankfurt am Main, Germany }
\address{$^2$Joint Institute for Nuclear Research, Dubna 141980, Russia}
\begin{abstract}
We calculated the single spin asymmetries for the reaction 
$P+P(\uparrow)\rightarrow l~\overline{l}+X$ 
in the framework of twist-3 QCD for HERA energies. 
The necessary imaginary phase is produced by the 
on-shell contribution of the quark propagator, while 
the long distance part is analogous to that providing the direct photon
asymmetry calculated by J. Qiu and G. Sterman. The asymmetry turns out to
be generally of the order percent. 
\end{abstract}
\end{frontmatter}
\begin{figure}[htbp]
\vspace{1cm}
\centerline{\psfig{figure=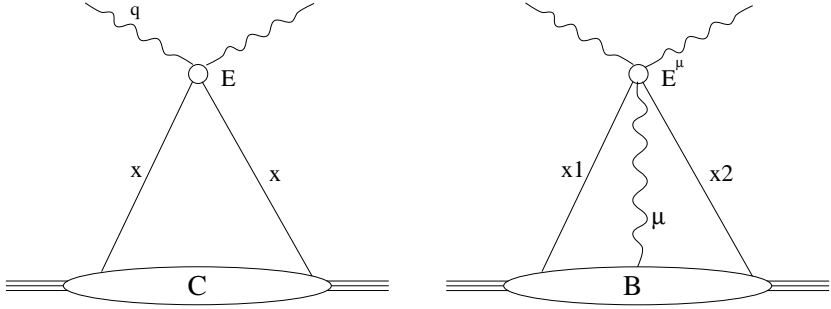,width=11cm}}
\caption[]{\it Parton distribution functions and parton subprocesses for twist 3.}
\label{fig1}
\end{figure}

Recently the study of single spin asymmetries has attracted great attention
[2-9] because 
they are relatively easy to measure e.g. at RHIC or 
at HERA (where, however, no suitable detector is presently available) 
and because 
single spin experiments promise to give very interesting information
on specific higher twist correlators and thus on the internal wave function 
of the nucleon.
The QCD description of these asymmetries is, however, 
rather complicated, due to the fact,
that there are different sources for the 
imaginary phases on the twist-3 level, namely, so-called fermionic and gluonic poles.
For the process discussed here 
a situation comparable to prompt photon production at $x_F <0$ appears:
a quark propagator is put on-shell, resulting in a soft gluon momentum, giving 
rise to a theoretically much cleaner situation.
One finds asymmetries in pion production to be of the order of ten 
percent or more \cite{ant}, but these asymmetries are still not 
understood at all on the QCD level and therefore could not be linked
so far to well defined properties of the nucleon. Other, theoretically
well defined, asymmetries are estimated to be rather on the 
percent level. A carefully  studied process is e.g.
prompt photon production \cite{gull}.
It is natural to ask how large the corresponding asymmetry would be 
for the closely related process of dilepton production. It turns out
that for this process in addition to graphs analogous to those for direct 
photon production there is another contribution which should dominate
the total cross section unless an associated quark-jet is required.
We find that the calculation of this new contribution can be done 
very precisely, based on first principles. The resulting asymmetry  
strongly depends on the kinematic domain.\\ 
A QCD analysis of the density matrix of a hadron \cite{Ter} gave for the partonic expression for the
spin dependent contribution to a hard cross section
\beqar{1}
W&=&\int dx~ C_{T}^{A}(x)\frac{1}{4}{\rm Tr}\left[ E(x)\lash S \gamma ^{5}\right]\nonumber \\
&+& \int dx_{1} dx_{2}~ B^{A}(x_{1},x_{2}) \frac{1}{4}{\rm Tr}\left[
E_{\mu} (x_{1},x_{2}) \lash p \gamma ^{5}\right]S^{\mu}\\
&+&\int dx_{1} dx_{2} ~ B^{V}(x_{1},x_{2}) \frac{1}{4}{\rm Tr}\left[ E_{\mu}(x_{1},x_{2})\gamma _{\rho}
\right] \epsilon ^{\rho \mu S p}\nonumber
\eeqar
where $\epsilon ^{\rho \mu S p}= \epsilon ^{\rho \mu \alpha \beta}~
S_{\alpha}p_{\beta}$ with $S_{\alpha}$ being
the covariant hadron polarization vector. $E(x)$ and $E^{\mu}(x)$ 
are the coefficient functions of the parton 
subprocesses with two quark lines and with two quark lines and one gluon line, 
respectively (fig.\ref{fig1}).
The generalized parton distributions are given by
\beqar{2}
C_{T}^{A}(x)&=& \int \frac{d\lambda}{2\pi} e^{i \lambda x}\times 
\bra{p,S}\overline {q}(0)\lash S \gamma ^{5} q (n\lambda)\ket{p,S}\\
\nonumber\\
B^{A}(x_{1},x_{2})&=&\int \frac{d\lambda _{1} d\lambda _{2}}{(2\pi)^{2}}
e^{i \lambda _{1}(x_{1}-x_{2})+i \lambda _{2}x_{2}}\nonumber \\
&&\times \bra{p,S}\overline {q}(0)\lash n \gamma ^{5} S\cdot D(n\lambda _{1}) 
q (n\lambda _{2})\ket{p,S}\\
\nonumber \\
B^{V}(x_{1},x_{2})&=&\int \frac{d\lambda _{1} d\lambda _{2}}{(2\pi)^{2}}
e^{i \lambda _{1}(x_{1}-x_{2})+i \lambda _{2}x_{2}} \nonumber\\
&&\times \bra{p,S}\overline {q}(0)\lash n  D_{\mu}(n\lambda _{1})q 
(n\lambda _{2})\ket{p,S}
\epsilon^{\mu S \overline{n} n}
\eeqar

\begin{figure}[htbp]
\vspace{1cm}
\centerline{\psfig{figure=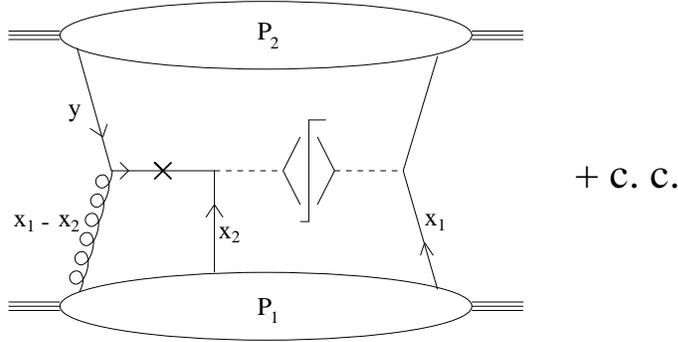,width=9cm}}
\caption[]{\it Twist 3 diagramm for the process $P+P(\uparrow)\rightarrow l~\overline{l}+X$. The quark propagator
generates the imaginary part when it is put on-shell, 
therefore the gluon has vanishing four-momentum 
$(x_{1}-x_{2})\approx 0$.}
\label{fig2}
\end{figure}

The matrix element $T$ introduced by Qiu and Sterman in \cite{qs}
is related to a piece of $B^V$  which is singular in $(x_1-x_2)$ 
by use of the equation of motion and the gauge condition:
\beqar{5}
T(x)=\lim_{x_{1}\to x_{2}} T(x_1,x_2)\nonumber\\
T(x_1,x_2)=\pi{{(x_1-x_2)}\over g} B^{V}(x_{1},x_{2})=
- \int \frac{dy_{1}^{-}}{4\pi}e^{ix_1 p^{+}y_{1}+iy_2 p^{+} (x_2 - x_1)}
\nonumber\\
\bra{p,S_{T}}\overline{q}(0) \gamma ^{+}
\int dy^{-}_{2} \epsilon _{\sigma \rho \alpha \beta}~S^{\sigma}_{T}~
n^{\alpha}~\overline{n}^{\beta}~
G^{\rho +}(y^{-}_{2})~
q (y^{-}_{1})\ket{p,S_{T}}
\eeqar
with the vectors $\overline{n}^{\beta}=
p_+ \delta _{\beta +}$ and $n^{\alpha}= \delta _{\alpha -}/p_+$.
This expression is of the same type as the 
correlators given in \cite{qs}. For the matrixelement we used a 
form suggested in \cite{qs}:
\beqar{6}
T(x)=const. \cdot F_{2}(x) {\rm GeV}.
\eeqar
The constant in front of the structure 
function $F_{2}(x)$ was estimated in \cite{bruno}
by connecting the matrix element to 
the second moment of the twist three part of the structure function 
$g_{2}(x)$. 
This yielded  for the relevant matrix element 
\beqar{8}
gT^{P}(x) = (0.08\pm 0.16)\cdot c \cdot F_{2}^{P} (x)~ {\rm GeV}
\eeqar
with $c$ in the range $1/3 < c < 1$. For our calculations we chose 
$c=1/3$ and for the constant in front of
$F_{2}^{P}(x)$ the value $0.08$.
The first term in (\ref{1}) does not contribute to the left-right asymmetry 
in the lowest order of $\alpha _{s}$, i.e. 
in the parton picture, because it has no imaginary part to compensate 
the imaginary trace of the Dirac matrices with $\gamma ^{5}$. 
The required imaginary part can only appear 
in one-loop approximation, so that any 
asymmetry generated by this term is proportional to $\alpha _{s}$.\\
In the case of the twist 3 subprocess, 
represented by the second and the third term in (\ref{1}), the situation is
different. 
In this case one has two imaginary parts; the first one is generated by 
cutting an internal quark propagator, i.e. putting it on
mass shell. 
The second one arises where 
an additional gluon is attached to a Born subprocess. 
The imaginary delta function contribution thus generated 
implies that the twist 3 term is the leading 
contribution to the single spin asymmetry under discussion.\\
We calculated the left-right asymmetry for the process 
$P+P(\uparrow)\rightarrow l~\overline{l}+X$ which is 
dominated by the Drell-Yan process
at small transverse momenta.
Left-right asymmetry means that the transverse momenta of the leptons
are correlated with the direction $\vec S\times \vec e_z$, where $S_{\mu}$
is the transverse polarization vector of the nucleon and $\vec e_z$ is the
beam direction. 
This process will be accessible in experiments 
with a polarized target at HERA \cite{WDN}
or with polarized beams at RHIC.
Note that the averaging over relative lepton
momenta gives a zero result because the three independent
vectors $S_{\mu}$, $P_{1\mu}$, and $P_{2\mu}$, where the latter two are the 
momenta of the incomming protons, are not 
enough to construct an invariant scalar containing
one epsilon tensor. 
The necessary fourth direction is given by the lepton angles, 
in complete analogy to the transversity studies
for the double transverse spin asymmetry. Thus the planned measurement
of the latter will also provide a measurement of the single spin 
asymmetry under consideration here. To extract it one just has to
average over the 
polarizations of one of the protons.  
The only twist 3 diagram contributing to the 
process we consider is shown in fig.\ref{fig2}. 
The on-shell propagation of the quark (indicated by a cross), results
in the gluon momentum to be almost zero, i.e. soft. 
In the cases treated by Qiu/Sterman \cite{qs} 
and Efremov/Teryaev \cite{Ter1} one encounters 
soft gluon poles when an internal gluon propagator is put on mass shell
and soft fermion poles when a quark propagator is put on mass shell.
In our case, however,  a quark propagator
is put on-shell resulting in a soft gluon momentum. 
A similar situation appears for prompt photon
production in the region of negative Feynman-x with $x_{F}=(t-u)/s$.
The use of Feynman rules leads to the following expression for the 
polarized part of the asymmetry
\beqar{AA}
\sigma_{pol.}&=&\frac{g}{\pi}\int dx_{2} ~\frac{T(x_{1},x_{2})}{(x_{1}-x_{2})}q(y)\nonumber\\
&&\cdot {\rm Tr}
\Bigg [ \lash P_{2} \gamma ^{\mu} \frac{(x_{1}-x_{2})\lash P_{1}+y\lash P_{2}}
{(x_{1}-x_{2})ys+i\epsilon}
\Delta \!\!  \lash l~ \gamma ^{\rho} ~\Delta  \!\! \lash l \Bigg ] 
\epsilon ^{\rho \mu S P_{1}}.
\eeqar
where $\Delta l^{\mu} =l_{1}^{\mu}-l_{2}^{\mu}$ and 
$l^{\mu}=l^{\mu}_{1}+l^{\mu}_{2}$.
The last factor in the bracket follows from 
$\lash l_1\gamma^{\rho} \lash l_2+
\lash l_2\gamma^{\rho} \lash l_1=
(\lash l\gamma^{\rho} \lash l-\Delta \lash l\gamma^{\rho} \Delta\lash l)/2$
and the fact that the vector  $\Delta l^{\mu}$ is needed for the contraction
with the epsilon tensor, such that only the last term contributes.\\
The additional imaginary part, discussed earlier, 
is yielded by the internal on mass shell quark propagator 
\beqar{DD}
{\rm Im} \frac{1}{(x_{1}-x_{2})ys+i\epsilon}=-i\pi \frac{1}{ys}\delta (x_{1}-x_{2}).
\eeqar
The mirror diagramm cancels the real part and gives a 
factor of two for the imaginary part leading to
\beqar{DD1}
&&\sigma _{pol.}= \\
&-&2ig \int dx_{2} ~T(x_{1},x_{2})q(y){\rm Tr}
\Bigg [ \lash P_{2} \gamma ^{\mu}\lash P_{1}\frac{1}{ys}\delta (x_{1}-x_{2})
\Delta  \!\! \lash l~ \gamma ^{\rho} ~\Delta \!\!  \lash l \Bigg ] \epsilon ^{\rho \mu S 
P_{1}}\nonumber\\
&-&2ig \int dx_{2} ~\frac{T(x_{1},x_{2})}{(x_{1}-x_{2})}q(y){\rm Tr}
\Bigg [ \lash P_{2} \gamma ^{\mu}\lash P_{2}y\frac{1}{ys}\delta (x_{1}-x_{2})
\Delta \!\!  \lash l~ \gamma ^{\rho} ~\Delta \!\!  \lash l \Bigg ] \epsilon ^{\rho \mu S 
P_{1}}\nonumber.
\eeqar
By using the Taylor expansion for the matrixelment up to first order
\beqar{EE}
T(x_{1},x_{2})=T(x_{1},x_{1})+\frac{\partial T(x_{1},x_{2})}{\partial x_{1}}\cdot 
(x_{1}-x_{2})
\eeqar
and by using $\Delta \!\! \lash l~ \gamma ^{\rho} ~\Delta \!\! \lash l=
2\Delta l^{\rho}~ \Delta \!\!  \lash l$ 
plus a term which vanishes when contracted with the epsilon tensor
one finally derives the following expression for the 
polarized part
\beqar{FF}
\sigma _{pol.}\sim 2g ~T(x_,x)\frac{q(y)}{ys}(P_{1}\cdot \Delta l)+
2g ~\frac{d T(x_,x)}{dx}\frac{q(y)}{s}(P_{2}\cdot \Delta l).
\eeqar
The matrixelement $T(x,x)$, that depends only on one momentum fraction, is given by the one 
introduced by Qiu and Sterman in \cite{qs} as shown in (\ref{5}).
For the normalization one has to treat the twist two part in the same way which yields 
\beqar{GG}
\sigma _{unpol.}\sim \left[ 2\left( P_{2} \cdot \Delta l\right)~\left( P_{1} \cdot \Delta l\right)
-2\left( P_{2} \cdot l\right)~\left( P_{1} \cdot l\right)\right]q(y)q(x)
\eeqar
with the unpolarized quark distribution $q(y)$.
The asymmetry finally reads
\beqar{4}
A=2g\frac{\left[ (P_{1}\cdot \Delta l) \frac{T(x,x)}{y}+(P_{2}\cdot \Delta l)
\frac{dT(x,x)}{dx} \right]}
{s \left[ \left( P_{1}\cdot  l\right)\left( P_{2}\cdot l\right) - \left( P_{1}
\cdot\Delta l\right)
\left( P_{2} \cdot\Delta l\right)\right] q(x)}~\epsilon^{P_{1}P_{2}\Delta l S}.
\eeqar
In the c.m. frame, this simplifies to  
\beqar{4a}
A=g~ \frac{\sin 2\theta ~\cos \phi \left[T(x,x)-x\frac{dT(x,x)}{dx}\right]}
{M\left [ 1+\cos^2 \theta  \right] q(x)}
\eeqar
with
\begin{eqnarray}
S_{\mu}&=& (0,1,0,0) \nonumber \\
l_{1\mu}&=&E_e(1,\sin\theta \sin \phi, 
\sin\theta\cos\phi, \cos \theta)\\
l_{2\mu}&=&E_e(1,-\sin\theta \sin \phi, 
-\sin\theta\cos\phi, -\cos \theta)\nonumber \\
\end{eqnarray}
in the dilepton c.m. system 
and with the dilepton mass $M$.
To take into account different quark flavours one
has to sum the numerator and denominator multiplied with the different antiquark density 
$\bar q(y)$ and weighted with $e_q^2$.
To illustrate the content of (\ref{4a}) we calculated A as a function of $\theta$.
We used the HERA energy of 820 GeV (fixed target) and chose different values for the 
momentum fraction of the scattered quark from the polarized proton, namely 
$x=0.012$, $x=0.1$ and $x=0.5$ with a fixed dileptonmass of $M^2 =10 GeV^2$.
We also chose $\phi = 0$, where the asymmetry is largest. In practical
experiments one may define the integrated
asymmetry by averaging over the angles with the weight sign($\theta \cos\phi)$
(otherwise, the angle integration 
will give zero result). 
The result for $x=0.5$ is shown in figure \ref{B}. For a value of $x=0.012$ one 
yields a value of $0.015\%$ for the asymmetry and for $x=0.1$ the asymmetry 
reaches $0.2\%$, showing the increase with the growth of the momentum fraction. 
\begin{figure}[htbp]
\centerline{\psfig{figure=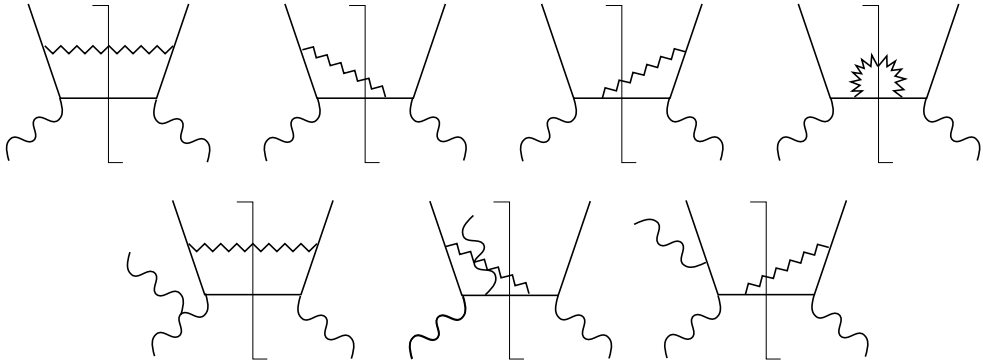,width=13cm}}
\caption[]{\it Typical twist-2 (upper row) 
and twist-3 (lower row) diagrams for the case 
of high $p_T$ dilepton production. The outgoing 
photon can produce a dilepton pair. This process
can be distinguished from the one treated here by the 
additional jet coming from the
fragmenting quark.
}
\label{qiu} 
\end{figure}
\begin{figure}[htbp]
\centerline{\psfig{figure=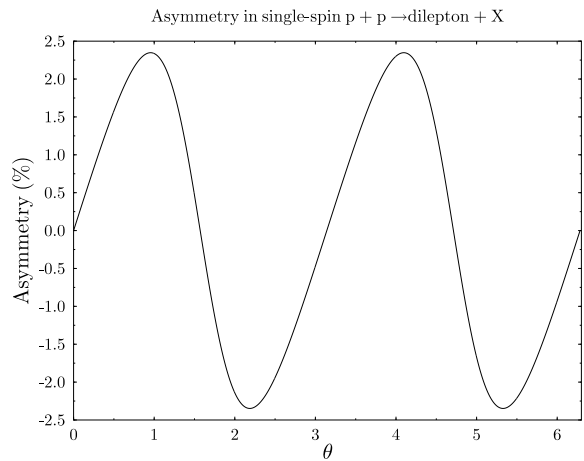,width=11cm}} 
\caption[]{\it Asymmetry for the process $P+P(\uparrow)\rightarrow l~\overline{l}+X$ 
with the momentum fraction chosen as $x=0.5$ at the HERA energy of 
820 GeV, $M^2 =10 GeV^2$.}
\label{B}
\end{figure}
The observable growth of the asymmetry at large $x$ due to the contribution of $dT/dx$
was first discussed in \cite{qs} for the photon case. 
In principle, the case treated here can be separated in the experiment from 
high transverse momentum dilepton production in the 
Qiu and Sterman approach (generalized, for the fermionic poles contribution, 
to the case of finite dilepton masses as well \cite{KT95}) 
since in our case one has only the dilepton and the two 
remaining hadrons. In the case of Qiu/Sterman and Efremov/Teryaev one has an additional jet coming 
from the fragmentation of an outgoing quark balancing the photon/dilepton 
momentum (see figure \ref{qiu}).
Also, because for high $p_T$ dileptons there are $2 \to 2$ subprocesses, 
the asymmetry is non-zero, when the averaging over lepton angles is performed, contrary to our case.
Nevertheless, it is not clear whether the resolution is high enough to really separate 
the two cases in a practical experiment. 
\\
We are indebted to W.-D. Nowak for numerous stimulating discussions of
single spin experiments.
O.T. is endepted to B. Geyer for warm hospitality at the Centre of Advanced Studies, 
University of Leipzig, where essential part of this work was performed. His work was 
partially supported by the Russian Foundation for Fundamental Research and INTAS.

\end{document}